# Controlling Terahertz Spintronic Photocurrents in 2D-Semiconductor|Ferromagnet Heterostructures through a Functional Hybrid Interface

Authors: A. Alostaz[1,2,*], R. Rouzegar[1], Eddie Harris-Lee[3,7], Xinhou Chen[4], Shijie Wang[5], Kuan Eng Johnson Goh[5,6,8], D. E. Bürgler[2], H. Yang[4], Elbert E. M. Chia[6], S. Sharma[1,7], T. Kampfrath[1], T. S. Seifert[1,*]

[1] Department of Physics, Freie Universität Berlin, Berlin, Germany

[2] Peter-Grünberg-Institute (PGI-6), Forschungszentrum Jülich, Jülich, Germany

[3] Max Planck Institute of Microstructure Physics, Weinberg 2, D-06120 Halle, Germany

[4] Department of Electrical and Computer Engineering, National University of Singapore. Singapore, Singapore

[5] Institute of Materials Research and Engineering, A*STAR (Agency for Science, Technology and Research), Singapore, Singapore

[6] Division of Physics and Applied Physics, School of Physical and Mathematical Sciences, Nanyang Technological University, Singapore 637371, Singapore

[7] Max-Born-Institute for Non-Linear optics, Max-Born Strasse 2A, Berlin, Germany

[8] Centre for quantum Technologies, National University of Singapore, 3 Science Drive 2, Singapore, 117543, Republic of Singapore

* Corresponding author: tom.seifert@fu-berlin.de, afnan.alostaz@fu-berlin.de

**Abstract**

A profound understanding of terahertz (THz) spin and charge currents in heterostructures involving ferromagnets (FMs) and two-dimensional (2D) materials promises emerging applications in high-speed sensing and data processing. Yet, ultrafast experimental insights remain very limited. Here, we study the efficient photo-generation of THz spin and charge currents in bilayers made from the transition-metal dichalcogenide (TMD) $MoS_2$ and the FM Co. We find that the efficiency of current generation strongly depends on the pump photon energy, as previously reported. Surprisingly, however, we observe that the current dynamics remain identical for pump photon energies above and below the $MoS_2$ band gap. Supported by *ab-initio* calculations, we conclude that an interfacial hybrid metallic layer forms at the $MoS_2$/Co boundary that has a pronounced photon-energy-dependent absorptance. Thus, the hybrid interfacial layer effectively acts like a pump-energy transducer that increases the spin-current generated in the nearby Co. Our results uncover the vital role of interfacial hybridization as a yet unexplored mechanism for efficient generation of ultrafast photocurrents in 2D-TMD|FM structures.

## Introduction

Interfacing two-dimensional (2D) materials with other compounds holds great promise for novel applications in data processing, catalysis, or solar-energy conversion [1, 2]. A particularly large potential lies in stacks made from ultrathin metals and 2D semiconductors because they are potentially compatible with existing CMOS technology [3-5] and allow one to exploit emerging functionalities, such as magnetism, superconductivity, multiferroicity or strong spin-orbit interaction [6-11], that can arise from proximity or hybridization effects close to interfaces [12-17].

As an example, nanostructures consisting of a ferromagnetic metal (FM) and a quasi-2D semiconductor (SC), like a transition-metal dichalcogenide (TMD), can combine spin- and valleytronic effects [2, 18-22]. Despite these promises, a long-standing roadblock for applications based on FM|SC stacks is the efficient injection of spins from the FM into the SC [23]. To bypass this efficiency barrier, strongly-out-of-equilibrium spin-current injection on ultrashort time scales was suggested [24, 25], which could, in addition, permit ultrafast device operation.

In this regard, previous works [26-31] studied light-induced ultrafast spin and charge transport in FM|TMD stacks by terahertz (THz)-emission spectroscopy considering solely amplitude changes of THz signals. More specifically, Cheng et al. [26] focused on bilayers made from the FM Co and the TMD $MoS_2$. Following femtosecond optical excitation, they found efficient THz-field generation that was ascribed to ultrafast spin-current injection from Co into the TMD. Within their interpretation, the absorption of pump photons generates electrons and holes at energies up to the photon energy away from the Fermi energy. Importantly, only spin-polarized photocarriers with sufficiently high energy could overcome the metal-semiconductor energy barrier. Consistent with that expectation, their experiment revealed a pronounced dependence of the measured terahertz (THz) photocurrent amplitude on the driving photon energy. Importantly, since highly nonthermal photocarriers in a metal typically relax within only a few tens of femtoseconds depending on the pump photon energy, this proposed ultrafast spin-current-injection scenario should also lead to distinct dynamic changes in the measured THz current. However, previous studies lacked the required time resolution to reveal such dynamical features.

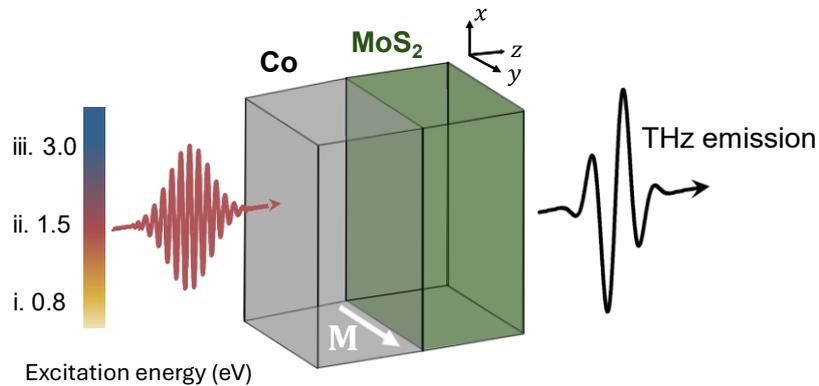

Figure 1. Experimental schematic. Ultrafast photocurrent spectroscopy of a $MoS_2$|Co heterostructure using, subsequently, three different pump photon energies of i. 0.8 eV, ii. 1.5 eV, and iii. 3.0 eV. An external magnetic field sets the sample magnetization M to be in the sample plane, while we record the emitted THz electric field polarized in the plane perpendicular to M.

Here, we test the previously proposed ultrafast spin-current-injection scenario from Co into $MoS_2$ by exciting $MoS_2$|Co bilayers with varying pump photon energy and measuring the pump-induced THz charge dynamics with ∼ 50 fs temporal resolution. As a fully metallic reference, we compare to Co|Pt [32, 33], where the pump pulse heats up the FM layer, leading to spin-current injection into the Pt layer in proportion to the ∼ 100 fs demagnetization dynamics of Co [34, 35]. Consequently, the inverse spin Hall effect converts the spin current inside Pt into a transverse charge current that radiates a THz pulse.

Strikingly, the THz emission from $MoS_2$|Co exhibits pump-induced current dynamics indistinguishable from those of Co|Pt, independent of whether the excitation energy exceeds the metal-semiconductor Schottky barriers. The identical temporal response indicates that the spin-current dynamics in $MoS_2$|Co are governed by the Co layer and are not measurably modified by

the presence of the TMD, challenging the previously proposed ultrafast spin-injection scenario [24-26].

Instead, we propose that the quasi-universal dynamics of the induced photocurrent combined with its marked pump-photon-energy dependence in terms of amplitude is a direct consequence of a hybrid layer forming at the $MoS_2$/Co interface, effectively transforming the semiconducting $MoS_2$ partially into a metal. This hybrid layer features pronounced changes in absorption coefficient with excitation energy. Our findings are supported by fully consistent *ab-initio* calculations, confirming the presence of strongly hybridized electronic wavefunctions in the $MoS_2$/Co interfacial region. These hybrid electronic states lead to a partially metallic and magnetic character of the hybrid layer reminiscent of Co, while simultaneously retaining signatures of the TMD semiconductor band structure, featuring a strong energy dependence of the absorption coefficient around the optical band gap. Following the enhanced pump absorption in the hybrid layer, the additional localized energy leads to an increased spin-current emission by the nearby Co region.

Such a hybrid layer has long been predicted theoretically [13-15, 22, 36, 37] and found in some experiments [16, 38] but its impact on ultrafast spin-current generation has not been studied experimentally. Thus, our study not only validates the long-standing paradigm of inefficient spin-current injection from metals into semiconductors [23] but also highlights new opportunities to tune THz-spin-current sources through hybrid-layer functionalization.

**Experiment**

As schematically depicted in Fig. 1, we use THz-emission spectroscopy to study the dynamics of ultrafast photocurrents in $MoS_2$|Co stacks for different pump-photon energies with ~ 50 fs time resolution. The detailed structure of the studied samples reads $Al_2O_3$(500 µm)|$MoS_2$(1 monolayer ≈ 0.7 nm)|Co(5 nm)|$SiO_2$(3 nm), literally the same sample used in Ref. [26], and $Al_2O_3$(500 µm)|Co(5 nm)|Pt(2 nm). The latter sample serves as a pump-photon-energy independent reference that is known to monitor the dynamics of the pump-induced excess magnetization in Co [32, 33]. By using such a reference, we can correct for setup-related modifications when changing the pump-photon energy. In our experiment, a linearly-polarized pump pulse excites the sample, resulting in the creation of an ultrafast photocurrent that radiates an electromagnetic pulse into the far field with frequencies extending into the THz frequency range. We focus on the dominant THz-electric-field component that is polarized perpendicular to the sample magnetization M. It is detected by electro-optic sampling [39] in a suitable detection crystal, resulting in a THz-emission signal $S$ vs time $t$ (see Supporting Information S1 for more details on the THz-emission setup and the sample preparation).

As we are interested in spin-related THz signals $S$, we reverse the Co in-plane magnetization M (Fig. 1) with an external magnetic field of about 20 mT and focus on THz-signal contributions odd under this operation, i.e., $S(t) = [S(t, +M) − S(t, −M)]/2$. We note that THz-signal contributions even in M, which contain charge-transfer processes across the interface, are found to be about one order of magnitude weaker (see Supporting Information S1.3).

Importantly, we choose three photon energies of 0.8, 1.5 or 3.0 eV, to excite the sample in distinctly different regimes [26], where (i) either only hot electrons (0.8 eV) or (ii) both hot electrons and hot holes (1.5 eV) could be potentially injected across the energetic barrier in between Co and $MoS_2$, or where (iii) direct excitation across the $MoS_2$ electronic band gap is possible (3.0 eV).

**Results and interpretation**

**Photocurrent signals.** Our main experimental results are summarized in Fig. 2. In detail, Fig. 2a presents the time-domain THz-emission signals $S(t)$ from MoS$_2$|Co and Co|Pt for different pump-photon energies. The waveforms from MoS$_2$|Co are normalized for each pump-photon energy separately such that the signal maxima from MoS$_2$|Co coincide with those of Co|Pt, resulting in the corresponding amplitude ratios $S_{\text{MoS}_2|\text{Co}}/S_{\text{Co|Pt}}$. Remarkably, across all three pump-photon energies, the THz-emission waveforms of MoS$_2$|Co have the absolutely same dynamics as the Co|Pt reference within the accuracy of the measurement. We note that changes in THz-signal waveforms for different pump photon energies are related to changes in the experimental setups, which can be accounted for by our Co|Pt reference sample, whose response is independent of the pump photon energy. According to Fig. 2a, variation of the pump photon energy induces changes only in the signal amplitude.

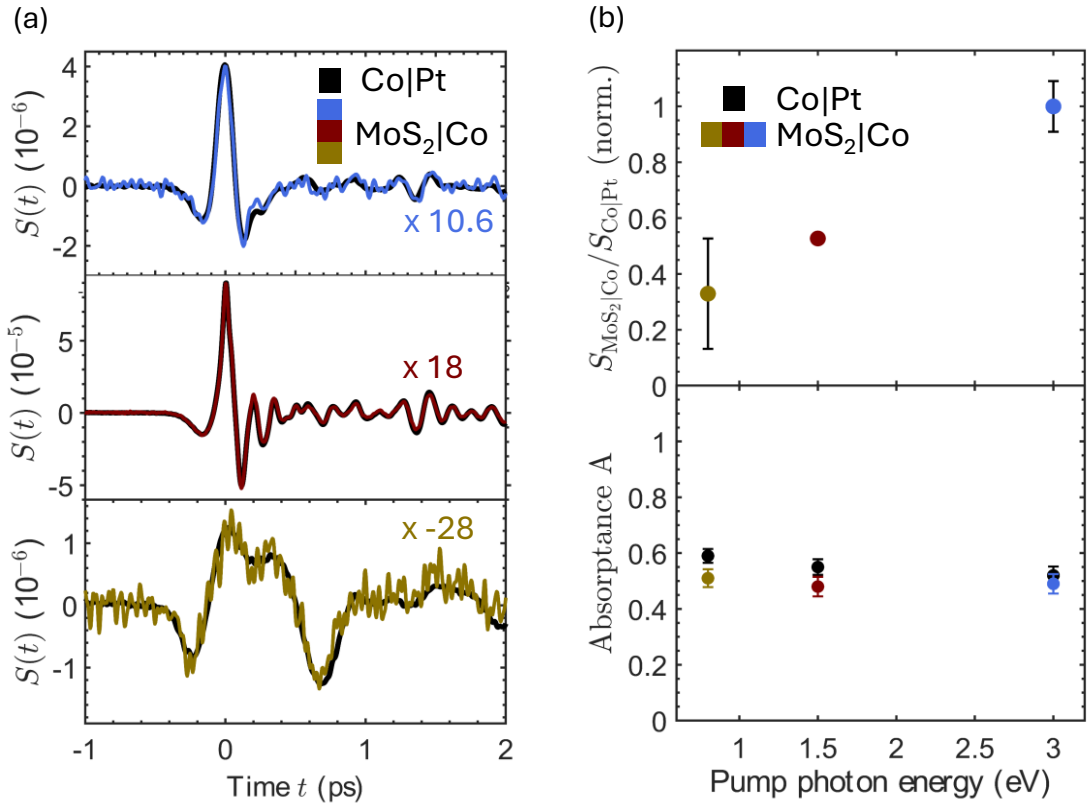

Figure 2. THz-emission signals from MoS$_2$|Co for different pump-photon energies. (a) THz-emission signals from MoS$_2$|Co and a Co|Pt reference sample for pump photon energies of 0.8 eV (dark yellow), 1.5 eV (dark red) and 3.0 eV (blue). Signal waveforms are scaled by the indicated factors, and waveforms from Co|Pt are multiplied by -1 for better comparison (b) Top: THz-emission amplitude ratios between MoS$_2$|Co and Co|Pt as a function of pump-photon energy. The amplitude ratios are all normalized by the value for 3.0 eV excitation. Bottom: Corresponding total pump-light absorptances in MoS$_2$|Co (colorful markers) and Co|Pt (black markers) as a function of pump-photon energy. Error bars are obtained from uncertainties in the signal waveforms shown in panel (a).

Consequently, in Fig. 2b, we display the corresponding amplitude ratios of $S_{\text{MoS}_2|\text{Co}}$ and the reference $S_{\text{Co|Pt}}$ vs pump-photon energy, normalized to its value at 3.0 eV excitation. We find a strong dependence of the scaling amplitude ratio on the pump-photon energy, leading to an enhancement of scaling factor of about 2 and 4 when going from 0.8 eV to 1.5 eV and from 0.8 eV to 3.0 eV excitation. Importantly, the scaling with pump-photon energy is totally uncorrelated with the overall pump-light absorptance at the respective energies (Fig. 2b). This result agrees with the trend reported by Cheng and coworkers [26]. However, the identical dynamics between MoS$_2$|Co and Co|Pt are unexpected.

**Photocurrent dynamics.** To better understand these two central observations - the surprising energy-independent dynamics and the pronounced energy dependence of the THz-emission amplitude - we first review prior knowledge of the fully metallic reference THz emitter Co|Pt [32, 33]. The laser-induced THz emission from Co|Pt typically involves three essential steps: (i) optical generation of an excess magnetization in the FM, (ii) spin injection, and (iii) spin-to-charge conversion in the heavy metal layer (Pt). In step (i), the pump pulse deposits energy in the electronic system of Co. This ultrafast electron-heating causes a pump-induced excess magnetization in Co [34, 35]. In step (ii), the excess of magnetization relaxes locally via spin flips or nonlocally via spin transport into an adjacent layer, such as Pt. In step (iii), spin-orbit interaction converts the injected spins current in Pt into a transverse charge current, typically via the inverse spin Hall effect. The latter generates an ultrashort charge-current burst that leads to the emission of a THz pulse.

Importantly, previous studies on fully metallic stacks, such as Co|Pt, revealed that the pump-photon energy is largely irrelevant to the resulting amplitude of spin-current injection and its THz dynamics [40, 41]. Instead, the dynamics of spin injection are entirely determined by temporal evolution of the Co excess magnetization as steps (ii) and (iii) essentially happen instantaneously within the reported experimental time resolutions of about 10 fs [34, 35]. With this in mind, the observed identical dynamics in Fig. 2a strongly indicate that the same driving force is present in $MoS_2$|Co and in the Co|Pt reference, namely, the pump-induced excess magnetization in Co dictates the THz-emission dynamics.

This conclusion challenges existing interpretations [24-26] that require highly nonthermal spin-polarized photocarriers to be injected from the metallic FM into the TMD, along with crossing an energy barrier. Such a nonthermal-carrier scenario is expected to affect not only the amplitude but, most importantly, the photocurrent dynamics, since highly excited photocarriers decay within a few 10 fs after excitation in the metal [42]. As a result, the photocurrent and, thus, the measured THz-signal waveform should exhibit faster dynamics compared to the ones in Co|Pt, because any possible transport channels for spin-polarized highly nonthermal carriers close more rapidly than the typical dynamics of the photo-induced excess magnetization in Co [35]. In contrast, the identical dynamics observed for $MoS_2$|Co and the Co|Pt reference for different pump-photon energies (Fig. 2a) leads us to conclude that contributions from spin transport mediated by highly nonthermal photocarriers are minor in our experiment.

**Photocurrent amplitude.** However, it remains unclear what mechanism causes the marked photon-energy dependence of the photocurrent amplitude in $MoS_2$|Co compared to Co|Pt (Fig. 2b). To answer this important question, we perform *ab-initio* band-structure calculations of the $MoS_2$/Co interface (Fig. 3, more details are provided in the Supporting Information S2). From our calculations, we find a clear hybridization of electronic wavefunctions in the $MoS_2$/Co interfacial region that results in the formation of partially spin-polarized in-gap states, i.e., a metallic character and a magnetization of $MoS_2$ on Co. In other words, a hybrid layer forms by wavefunction overlap that contains a mix of Co and $MoS_2$ properties, fully consistent with literature reports in similar structures [13-17, 22, 37, 38].

To compare these calculations to our experimental results, we compute the pump absorption in the multilayer sample. We first determine the complex-valued electric-field amplitude $E$ of the pump pulse inside the sample by the thin-film approximation [43],

$$E = E_{\text{in}} \frac{2n_2}{1 + n_2 + Z_0 \sum_i d_i \sigma_i},  \qquad \text{Eq. 1}$$

with the incident electric field $E_\text{in}$, the substrate refractive index $n_2$, the free-space impedance $Z_0 \approx 377\,\Omega$, the film thickness $d_i$ of layer $i$ ($i = 1,2,3$ for Co, hybrid layer and MoS$_2$, respectively) and conductivity $\sigma_i$ at the pump-pulse center frequency. Second, we relate the electric field to the local absorptance $A_i$ in each layer $i$ via

$$A_i \propto d_i \text{Re}|E|^2 \,\text{Re}\,\sigma_i. \qquad \text{Eq. 2}$$

We find remarkable agreement between simulations based on Eqs. 1 and 2 (see Supporting Information S3) and our experiment for the hybrid-layer absorptance (Fig. 3c) and the total stack absorptance (Fig. 3d). The used optical conductivities $\sigma_i$ are taken from literature for Co and MoS$_2$ [44, 45], and from our *ab-initio* calculations for the Co-MoS$_2$ hybrid layer, assuming a hybrid-layer thickness of 1.4 nm. Thus, the hybrid layer extends over the entire MoS$_2$ monolayer (0.7 nm) and an equally thick part of the Co layer, consistent with our band-structure calculations (Fig. 4 and Supporting Information S2).

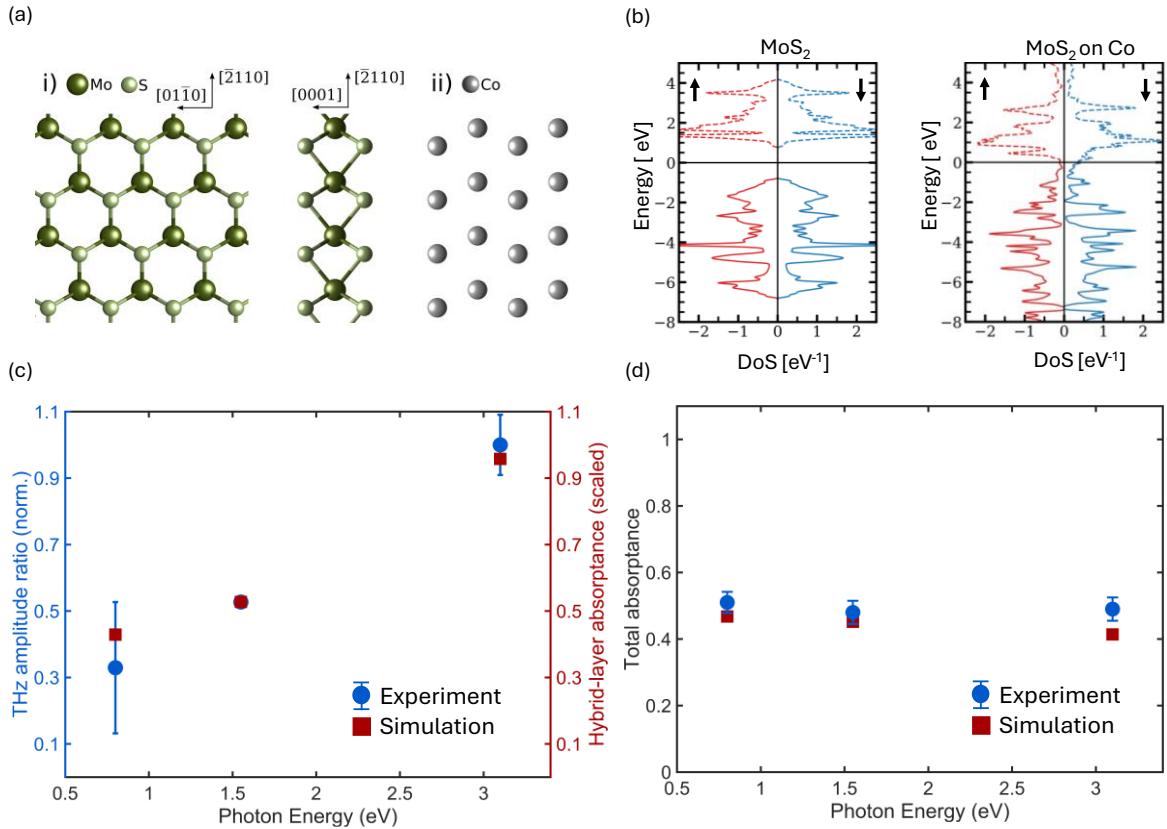

Figure 3. First-principles calculations of the MoS$_2$/Co interface and results of the constrained optical fitting model. (a) Atomic structures of a MoS$_2$ monolayer shown in-plane view (left panel) and along the c-axis (middle), with Mo atoms in dark green and S atoms in light green, and the Co lattice visualized along the c-axis (right, grey spheres). (b) Spin-resolved electronic density of states of an isolated MoS$_2$ layer (left) and of the interfacial layer of a monolayer of MoS$_2$ in contact with 5 monolayers of Co (right). (c) Photon-energy dependence of the hybrid-layer absorptance obtained from the multilayer-optical-absorptance model (red squares), compared to the experimentally measured signal ratios (blue circles with error bars, cf. Fig. 2b). A global scaling factor is applied to the simulated hybrid absorptance to account for the unknown absolute signal amplitude. (d) Corresponding simulated and experimental total absorptance $\sum A_i$ of the full sample stack (Sapphire|MoS$_2$|Co|SiO$_2$).

**Suggested scenario.** The excellent agreement between experiment and model calculations points to the following scenario (Fig. 4). The pump pulse is partially absorbed within the MoS$_2$|Co stack. Upon varying the pump-photon energy, the relative amount of absorbed energy within the hybrid layer changes strongly (Fig. 3c), whereas the absorption in the Co layer remains comparatively constant. Thus, we propose that the hybrid interfacial layer acts like an energy transducer that supplies additional ultrafast electron heating of the interfacial Co, thereby leading to an enhanced generation of excess magnetization therein. In turn, the excess magnetization can relax via spin injection toward the interface to the hybrid layer, where spin-to-charge conversion leads to the emission of a THz pulse [33].

Following our proposed scenario of an enhanced hybrid layer heating, it might appear surprising that the excess energy remains contained within the surface-near Co layers. One could naively expect that electrons quickly traverse the entire metallic sample region and distribute the energy homogeneously therein. This consideration suggests that the excited electronic states from the hybrid layer transfer energy primarily to states in the Co that are less delocalized, such as d-like electronic states, possibly related to magnon generation [34]. Such a scenario would be fully consistent with the derived dominant d-like character of the electronic states originating from Mo in the former gap region of the hybrid layer (Fig. S2.2) and the strong hybridization of these Mo with Co states across the interface (Fig. S3).

The predominant d-state-like electronic character would further rule out a dominant contribution of the inverse spin Hall effect to the THz charge current as it prevents efficient electron-mediated spin transport. Consequently, spin-charge conversion most likely arises from an inverse Rashba-Edelstein effect. The spin relaxation time of the related states is significantly shorter than our time resolution of ∼ 50 fs, similar to the Pt layer in the Co|Pt reference stack.

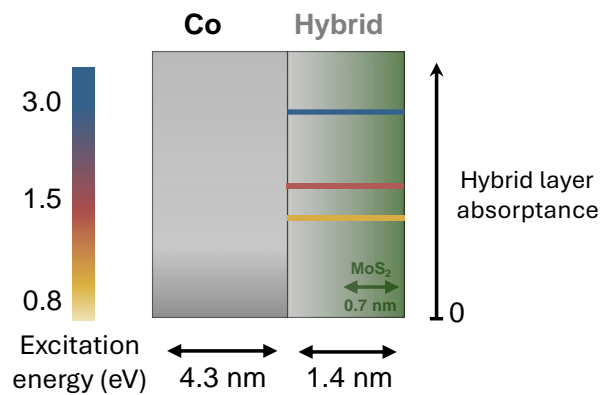

Figure 4. Schematic illustration of spintronic photocurrent generation in the hybridized Co|MoS$_2$ stack. The hybrid layer effectively acts as an energy transducer with strongly pump-photon-energy-dependent absorptance. When increasing the pump-photon energy, the excessively heated hybrid-layer electrons additionally increase the temperature of the electrons in the nearby Co layers, leading to an increased excess Co magnetization. Thus, the interfacial Co layers inject more spins into the hybrid interfacial region, where the resulting spin accumulation creates an ultrafast charge-current burst via spin-orbit coupling, giving rise to the emission of a THz pulse.

**Discussion.** With our understanding, we can exclude possible alternative photocurrent-generation mechanisms as follows. In the first alternative scenario, the hybrid layer itself develops a light-induced excess magnetization that significantly contributes to the resulting spin dynamics. However, we would like to note that the precise dynamics of the excess magnetization are defined by the specific material parameters describing the coupling between spins, electrons and the lattice. Therefore, identical dynamics observed for MoS$_2$|Co and Co|Pt strongly suggest that the dominant material defining the spin dynamics is Co and not the hybrid layer. A possible reason why the hybrid layer plays a minor role in this sense is the relatively small density of states (Fig. 3b) close to the Fermi energy compared to Co and the spin moment per atom that is around 1% of that of the unhybridized Co.

In the second alternative scenario, no hybrid layer is considered, and the pump pulse generates both an excess magnetization in Co and electron-hole pairs in the TMD. Only after these free carriers in the TMD are generated, spins can be efficiently injected into the TMD. Otherwise, charge backflow would be prohibited, resulting in charge accumulation and strongly reduced spin transport. Note that this scenario would result in a square-like pump-fluence dependence to provide both the excess magnetization in Co and the electron-hole pairs in the TMD. However, in our experiment, we find a linear pump fluence dependence (see Supporting Information S5.1), consistent with previous work [26].

The third alternative scenario is a slight variation of the second one. Here, the quadratic fluence dependence is absent because one of the two processes (generation of excess magnetization in Co or electron-hole generation in the TMD) saturates. A possible pathway to saturation would be an accumulation of electron-hole pairs over many laser shots, provided they were longer-lived than 12.5 ns, i.e., time between two consecutive laser pulses. We test this scenario by the generation of electron-hole pairs in the TMD with a 3.0 eV pump pulse in addition to a 1.5 eV pump. The resulting THz emission is found to be identical to the one without the additional 3.0 eV pump (see Supporting Information S6.1), thereby rendering the proposed third alternative scenario as unlikely.

We note that our proposed hybridization scenario stands out because the interfacial hybrid layer plays a crucial role in driving the THz current. This notion differs strongly from other THz-emission reports [46, 47] where an interfacial alloying layers forms and acts as THz spin-to-charge converter. While the discussions here are focused on the formation of a hybrid layer at the $MoS_2$/Co interface, the complex nature of TMD/FM contacts can lead to a marked lateral heterogeneity [48-50], which significantly impacts the local interfacial properties. This feature makes it challenging to probe the hybridization by spatially integrating probes, such as optical photoluminescence.

**Conclusion**

In summary, our findings strongly suggest that the semiconducting nature of the TMD $MoS_2$ has no direct impact on the THz spin-current dynamics in $MoS_2$|Co. Rather, an interfacial metallic hybrid layer forms at the $MoS_2$/Co interface, which dictates the marked dependence of spin-current amplitudes on the pump-photon energy. These results highlight the key role of interfacial hybridization in shaping ultrafast spin-current generation and facilitating its efficient conversion into charge currents in 2D-TMD|FM heterostructures. Beyond the specific $MoS_2$|Co system, the hybridization-mediated mechanism is likely ubiquitous in metal|semiconductor interfaces and may provide a universal blueprint for tailoring ultrafast spin-current generation through interfacial band-structure engineering.

**Acknowledgments**

The authors thank Chihun In and Yannic Behovits for fruitful discussions. The first phase of this study was supported by the Federal Ministry of Education and Research of Germany in the framework of the Palestinian-German Science Bridge (BMBF Grant No. 01DH16027). Open-access funding is enabled and organized by Projekt DEAL. H.Y. is supported by the Ministry of Education Singapore (MOE-T2EP50124-0006 and A-8003553-00-00).

E.E.M.C. acknowledges support from the Singapore Ministry of Education Singapore (AcRF Tier 3 Programme "Quantum Geometric Advantage" (Grant No. MOE-T2EP50124-0006MOET32023-0003), and A-8003553-00-00).AcRF Tier 2 "Dynamical characterization of the spin-to-charge conversion mechanisms in 2D heterostructures" (Grant No. T2EP50222-0047). S. Sharma would like to thank Leibniz Professorinnen Program (SAW P118/2021) and German Excellence Strategy – EXC3112/1 –533767171 (Center for Chiral Electronics) for funding. Computations were performed on the HPC System Viper at the Max Planck Computing and Data Facility. We gratefully acknowledge financial support from the German research foundation (DFG) through the Collaborative Research Center "Ultrafast spin dynamics" (Project No. 328545488, projects A04, A05, B02) and the Excellence Cluster EXC 3112 "Center for Chiral Electronics" (EXC 3112/1, Project No. 533767171), and the European Research Council through the ERC-2023 Advanced Grant ORBITERA (grant no. 101142285)